\documentclass[aps,twocolumn,pra,showpacs,floatfix]{revtex4}
\usepackage{epsfig}
\usepackage{graphicx}
\usepackage{dcolumn}
\usepackage{amssymb,amsmath}

\begin{document}

\newcommand{\Z}{Z_{\rm eff}}
\newcommand{\Zr}{\Z^{(\rm res)}}
\newcommand{\eps}{\varepsilon}
\newcommand{\eb}{\varepsilon _b}

\title{Atoms which can bind positrons}

\author{ C. Harabati, V. A. Dzuba, and V. V. Flambaum}

\affiliation{School of Physics, University of New South Wales, Sydney 2052,
Australia}

\begin{abstract}
Calculations of the positron binding energies to all atoms in the periodic table are presented and atoms where the positron-atom binding actually exists are identified. The results of these calculations and accurate calculations of other authors (which existed for several atoms only) are used to evaluate recommended values of the positron binding energies to the ground states of atoms. We also present the recommended energies of the positron excited bound levels and  resonances (due to the binding of positron to excited states of atoms) which can not emit positronium and have relatively narrow widths. Such resonances in positron annihilation and scattering  may be used to measure the positron binding energy.
\end{abstract}


\maketitle

\section{Introduction}

In this work, we apply the relativistic linearized coupled-cluster 
single-double (SD) approximation to calculate positron binding energies for the atoms in the whole periodic table. Calculating the positron-atom bound states is a challenging theoretical problem due to the strong electron-positron correlation effects and virtual positronium (Ps) formation~\cite{Massey64,Amusia76,Dzuba93}. 
The existence of such states was predicted by many-body calculation ~\cite{Dzuba95} and verified variationally~\cite{RM97,SC98} more than a decade ago. Since that time a number of theoretical papers were published but only few atomic systems were studied. The most accurate calculations were performed for eleven positron-atom systems involving Li, Na, Ag, Cu, Au, Be, Mg, Ca, Zn, Sr and Cd atoms~\cite{DFG99,BM06,BM10,M10,BP13,MZB08,DFH00,RM98,RM9844,MBR01,RMV98,M04}.  
Recent empirical fitted expression involving the polarizabilities ($\alpha$), ionization potentials ($I$), and numbers of valence $s$ electrons  has also been based on the best calculations mentioned above \cite{Babikov}.
A number of positron-atom bound states involving atoms with open $d$ subshells were studied in our previous paper~\cite{DFGH12}. In spite of all these predictions no experimental evidence for positron-atom bound states has been found so far.

The situation is better for positron-molecule interaction since the  resonant annihilation is observed for positrons in many polyatomic molecules \cite{GYS10}. The incident positron is captured into
the bound state with the target molecule, with the excess energy being transferred to vibrations. Since the vibrational motion
of the molecules is quantized, these transitions can only take place at
specific positron energies. These energies correspond to vibrational Feshbach resonances of the positron-molecule complex \cite{GYS10,Gr00,Gr01}. The majority of the resonances observed are associated with individual vibrational modes of the molecule. The energy of the positron binding  is then extracted from the downshift of the resonance energy,
relative to the energy of the vibrational excitation \cite{GBS02,BGS03}. Hence, by observing the resonances, the positron binding energy can be found. In this way, binding energies for
over 60 polyatomic species have been determined
\cite{DYS09,DGS10,JDGNS12,DJGNS12} by measuring positron annihilation using a high-resolution, tunable, trap-based positron beam \cite{GKG97}. 

A similar effect in atoms has been proposed for experimental detection of positron-atom bindings  in \cite{DFG10}.
It was suggested that the resonances in the positron annihilation with atoms can be observed and associated with binding of positron to a low-energy electronic excitations. These resonances can be found in open-shell atoms. If such atoms can bind a positron in the ground state, then it is very likely that they can bind a positron in the excited state of the same configuration.
One can then consider the following process,
\begin{equation}
e^+ + A\rightarrow e^+A^* \rightarrow A^+ + 2\gamma .
\label{eq:Ae+}
\end{equation}
First, a positron loses some energy by exciting the atom and becomes trapped to a bound state with the excited atom. 
Then, it annihilates with
one of the electrons, and the resulting gamma quanta can be detected. 
The first step of the process (\ref{eq:Ae+}) is obviously reversible. Hence, to estimate the efficiency of the resonant annihilation one needs to evaluate the rates of both positron annihilation ($\Gamma^a_\nu$) and autodetachment ($\Gamma^e_\nu$). One may estimate $\Gamma^e_\nu\sim 1-10$ meV for a Feshbach resonance at $ \varepsilon \sim 1$ eV, populated through a  quadrupole transition~\cite{DFG10}. Hence, the  resonances are sufficiently narrow to produce observable sharp features in the energy dependence of the annihilation rate $Z_{\rm eff}$. For a binding energy $ \eb=150$ meV, the estimated annihilation width is $\Gamma^a_\nu=4 \times 10^{-7}$ eV and  the branching ratio  $ \Gamma^e_\nu/\Gamma_\nu \approx 1$ ($\Gamma_\nu$ is the total width of the resonance). For a positron beam with the energy spread $ \delta \varepsilon \sim 25 $ meV, the peak resonant value of the annihilation rate is given by $Z_{\rm eff} \sim 10^3$ in Ref. \cite{DFG10}. This indicates that the positron-atom resonances could be observed with a trap-based-beam-technique similar to what was used for measuring resonances in the positron-molecule annihilation \cite{BGS03}. Resonances also manifest themselves in the positron scattering. 
  Another method - measurement of the positron binding energies through laser assisted photorecombination - was suggested in \cite{Surko}.

We also would like to mention a possibility to capture positron to a shallow bound level using  a pulse of a very strong magnetic field.
Such field of the strength up to 100 Tesla is available, for example, in the Los Alamos laboratory. Indeed, energy of an upper Zeeman component of the shallow positron-atom bound state may come above the ionization threshold and cross with the level of free positron resulting in the positron capture (the same  mechanism may be used to capture electron to a negative ion state). This possibility deserves a separate publication, and we do not proceed any further in the present paper.
 
These possibilities to create the positron-atom bound states motivate us to survey the whole periodic table  for positron binding and tabulate the results for experimentalists.

In present paper we extend the study started in our previous paper~\cite{DFGH12} to all atoms in the periodic table up to uranium. 
Almost all previous calculations considered positron interacting with either a closed-subshell atom or an atom with a single electron above a closed-subshell core. The only exceptions are our recent works~\cite{DFGH12,DFG10}. The reason for this is simple, there is no adequate theoretical method to perform accurate calculations for positron binding to open-shell atoms. In our previous paper\cite{DFGH12} we suggested to use the linearized coupled-cluster single-double (SD) approach for this purpose.
In this approach the interparticle interaction is included to all orders via an iterative procedure. The corresponding subset of terms includes the so-called {\em ladder} diagrams.
 This class of diagrams is very important in the positron-atom problem since it describes the effect of a virtual Ps formation. Summation of the electron-positron ladder-diagram series was performed earlier by solving a linear matrix equation for the electron-positron vertex function for hydrogen \cite{GL04}, noble-gas atoms~\cite{Ludlow04}, and halogen negative ions~\cite{LG10}.

The linearized coupled-cluster method in its single-double approximation has been applied for a number of very accurate calculations for atoms and ions with one external electron above closed shells (see, e.g., \cite{SD,BJS91,E3extra,SJD99,DJ07}). Hence it is expected that the
modified SD equations for the case of a positron interacting with a closed-shell atom should also give a reliable and accurate result.

A very brief summary of the SD equations for positron-atom interaction is provided in Section \ref{sec:theory}. The details of the
theory are given in our previous paper \cite{DFGH12}. Comparison of  our results for closed shell atoms with the available most accurate calculations is described in Sec. \ref{sec:closed}. This provides us with an  estimate of our accuracy. How the method is applied to the open-shell atoms is explained in Sec. \ref{sec:open} where we also present positron binding energies to the ground state configuration of every atom in the periodic table. Determination of the energies of the resonances and excited  bound states is explained in Sec. \ref{sec:excited}. The calculations for the positron binding to the excited electronic configurations of atoms are also presented in the same section. The paper ends with section \ref{sec:conclusion} where all our results are summarized.   

  The recommended values of the positron binding energies are presented in Table \ref{t:5}.  The recommended values of the excited bound states and resonances are presented in Tables \ref{t:6}, \ref{t:7} and \ref{t:10}. 

\section{Theory}\label{sec:theory} 

Many-body atomic calculations for the positron-atom binding need construction of the single-particle basis sets separately for electron states and positron states. We use relativistic Hartree-Fock (RHF) method and the B-spline technique~\cite{Bspline} to do this. The self-consistent RHF procedure is initially done for the atom without a positron. Then full sets of single-electron and positron states are constructed using B-splines in a cavity of a reasonably large radius $R$. The radius must be larger than the size of the atom and should be chosen in such a way that the total positron-atom system fits into the cavity. We use $R=30$ a.u. The effect of the finite cavity size on the positron-atom binding energy was studied and found to be negligible.

The single-particle basis states are found by constructing them as linear combination of B-splines and diagonalizing the matrix of the RHF Hamiltonian 
\begin{equation}
h = c {\bf \alpha}\cdot{\bf p}+(\beta-1)mc^2-\gamma \frac{Ze^2}{r}+\gamma V_{\rm d} - \zeta V_{\rm exch}.
\label{eq:dhf}
\end{equation}
Here, ${\bf \alpha}$, $ \beta $ are the Dirac matrices, $V_{\rm d}$ and $V_{\rm exch} $ are the direct and exchange RHF potentials respectively. The pair $(\gamma,\zeta)$ is taken $(1,1)$ for electron and $(-1,0)$ for positron. 
The wave function of an atom with a positron in state $v$ can be
written in the single-double (SD) approximation as an expansion
\begin{align}
|\Psi_{v}\rangle = &\left[ 1+\sum_{na}\rho_{na} a_n^\dagger a_a +
  \frac{1}{2}\sum_{mnab} \rho_{mnab} a_m^\dagger a^\dagger _n a_a a_b
   \right. \nonumber \\
&+\left. \sum_{r\neq v} p_{rv} a^\dagger _{r} a_{v}+ \sum_{rna}
  p_{rnva} a^\dagger_{r} a_{v} a_n^\dagger a_a\right]
|\Phi_{v}\rangle, \label{eq:psiv} 
\end{align}
where $|\Phi_{v}\rangle$ is the zeroth-order wave function of the
frozen-core atom in the relativistic Hartree-Fock approximation with
the positron in state $v$. It can be written as
\begin{equation}
|\Phi_{v}\rangle = a^\dagger_{v}|0_C\rangle,
\label{eq:0C}
\end{equation}
where $|0_C\rangle$ is the RHF wave function of the atomic core. Note that the following notations have been used to label the basis state in the rest of the paper: indices $a,b,c$ refer to electron states in the core, indices $m,n,k,l$ refer to electron states above the core, indices $v,r,w$ refer to positron states, and indices $i,j$ refer to any states. The expansion coefficients $\rho_{na}$ and $\rho_{mnab}$ in Eq.~(\ref{eq:psiv}) represent single- and double-electron excitations from the core. The coefficients $p_{rv}$ represent excitations of the positron, and the coefficients $p_{rnwa}$ represent simultaneous excitations of the positron and one of the electrons. The SD equations for the core excitation coefficients ($\rho_{na}$ and $\rho_{mnab}$) do not depend on the
external particle and they are the same in the electron and positron cases.
These are well known equations from the linearized coupled-cluster theory, the details of the theory can be found, for instance, in Ref.~\cite{SD}.
The first step is to solve these equations iteratively to obtain the single-electron coefficients $\rho_{ma}$ and the double-electron coefficients $\rho_{mnab}$ for the core and to fix them in the rest of the calculation. The convergence of the core equations is maintained by observing the correlation correction to the energy of the core. One can refer to our previous paper ~\cite{DFGH12} for the explicit form of the core equations. 

After solving the SD equations for the core, one can start iterating the SD
equations for the external particle. The SD equations for the positron can be obtained by substituting the state $|\Psi_{v}\rangle$ from Eq.~(\ref{eq:psiv})
into the relativistic many-body Schr\"{o}dinger equation,
\begin{equation}\label{eq:S}
H|\Psi_{v}\rangle = \epsilon_0 |\Psi_{v}\rangle.
\end{equation}
Projecting this equation onto $a^\dagger_w|0_C\rangle$
gives the equation for $p_{wv}$,
\begin{equation}\label{eq:v1}
(\epsilon_0-\epsilon_w)p_{wv} = 
 -\sum_{bm}q_{wbvm}\rho_{mb} + \sum_{bmr}q_{wbrm}p_{rmvb},
\end{equation}
Projecting Eq.~(\ref{eq:S}) onto $a^\dagger_{w}a^\dagger_{n}a_a|0_C\rangle$ gives
the equation for the double-excitation coefficient $p_{wnva}$,
\begin{equation}\label{eq:v2}
\begin{split}
(\epsilon_0+\epsilon_a -\epsilon_w-&\epsilon_n)p_{wnva} = q_{wnva} \\
-&\sum_{rm}q_{wnrm}p_{rmva} + \sum_m q_{wnvm}\rho_{ma}\\
-&\sum_b q_{wavb}\rho_{nb} + \sum_{mb}p_{wmvb}\tilde{g}_{mabn}\\
+&\sum_{rb}q_{warb}p_{rbvn} + \sum_{mb}q_{wmvb}\tilde{\rho}_{mabn}\, .
\end{split}
\end{equation}
In these equations $\tilde{g}_{mnkl} \equiv g_{mnkl} - g_{mnlk}$ and
$\tilde{\rho}_{mnkl} \equiv \rho_{mnkl} - \rho_{mnlk}$. The coefficients $g_{mnkl}$ and $q_{wnva}$ are the Coulomb integrals for the electron-electron interaction and electron-positron interaction respectively. When solving these equations, the correction to the energy of the positron state $v$,
\begin{equation}\label{eq:dev}
\delta \epsilon_v = - \sum_{mb} q_{vbvm}\rho_{mb} + \sum_{bmr}q_{vbrm}p_{rmvb},
\end{equation}
is used to control the  convergence.

In contrast to the electrons-only case the calculations do not stop here. When the SD equations are used to calculate the energy and the wave function of the atom with single-valence electron above closed shells the RHF approximation is already a good approximation for the valence electron and only small correction is needed. The correction is given by expressions similar to (\ref{eq:v1},\ref{eq:v2},\ref{eq:dev}) (see, e.g.~\cite{SD}). 
In the positron case there is no good zeroth-order approximation for the wave function of the bound positron. In the RHF approximation the positron-atom
interaction is repulsive, and all of the single-particle positron
basis states lie in the continuum. Since we cannot use a single positron RHF state as initial approximation we have to use all of them as a basis. The wave function of the positron bound to an atom is presented as a linear combination of the positron RHF states
\begin{equation}
\psi_p = \sum_v c_v \psi_v .
\label{eq:psip}
\end{equation}
The energy $\epsilon_0$ and the expansion
coefficients $c_v$ are found by solving the eigenvalue problem
\begin{equation}
\hat \Sigma X = \epsilon_0 X,
\label{eq:sx}
\end{equation}
where $X$ is the vector of expansion coefficients $c_v$, $\epsilon _0$ is the
lowest eigenvalue (which must be negative), and the elements of the effective
Hamiltonian matrix $\hat \Sigma$ are given by
\begin{equation}\label{eq:sigma}
\sigma_{vw} = \epsilon_v\delta_{vw} - \sum_{mb} q_{wbvm}\rho_{mb} +
\sum_{bmr}q_{wbrm}p_{rmvb} .
\end{equation}
The first term on the right-hand side of Eq.~(\ref{eq:sigma}) represents
the positron energies in the static RHF approximation. The second and third
terms describe the effect of the electron-positron correlations.
The SD equations (\ref{eq:v1}) and (\ref{eq:v2}) must be iterated for every state in the expansion (\ref{eq:psip}). 
Since these equations  depend on the
energy $\epsilon_0$ which is found later from Eq.~(\ref{eq:sx}), we start
with an initial guess for $\epsilon_0$. The calculations are then performed
iteratively, solving the SD equations (\ref{eq:v1}) and (\ref{eq:v2}) and
diagonalizing the matrix (\ref{eq:sigma}) several times until $\epsilon _0$
has converged.

The virtual Ps formation is described by electron-positron ladder diagram series. They are included in SD equation in all orders. However, some third order diagrams are missed by SD method. It is well known that the missing third-order terms can give sizeable corrections to the energy in atomic systems (see, e.g., Ref.~\cite{E3extra}). Including these terms can lead to significant improvements in the accuracy of the results (see the section \ref{sec:result}). Consequently, we include these contributions for the positron-bound states with atoms in this work. The list of the missing third-order diagrams in SD equations and the corresponding perturbation-theory corrections to the energy of the positron state are derived and listed in Ref.~\cite{DFGH12}.

\section{Results and discussion}\label{sec:result}

In our previous paper~\cite{DFGH12}, we have reported the calculations of positron binding to 26 neutral atoms by using the current method. The rest of atoms in the periodic table is examined in this paper up to atomic number $Z=92$ (uranium). All raw data are presented in Tables \ref{t:1} to \ref{t:4}. Although there is no rigorous criterion for the  positron binding to a neutral atom, it is widely accepted that the static dipole polarizability $ \alpha $ and  ionization potential $I$ of the atom play an important role. Indeed, positron feels strong attractive polarization potential $-\alpha/2r^4$ outside the atom. Therefore,  
 large value of $ \alpha$ and small radius of the atomic core increase the binding. Small radius of the  atomic core corresponds to a large ionization potential $I$. 
Moreover, $I$ and $\alpha$ may be combined as a single parameter which we have called the strength of the polarization potential ~\cite{Dzuba95}. It is a simple dimensionless parameter 
\begin{equation}
S= \frac{m \alpha I^2}{2\hslash^2}
\end{equation}
The ionization potentials $I$ from Ref.~\cite{NIST} and the static dipole polarizabilities $ \alpha $ from Ref.~\cite{Miller} are shown in Tables \ref{t:1}$-$\ref{t:4} for every atom in their ground state.
In Tables \ref{t:1}$-$\ref{t:4} we also present the values of the strength parameter $S$. 

It seems natural to classify all atoms up to $Z=92$ according to their ground state configurations, since  they have similar positron binding for  similar valence configurations. The arrangements of the first four tables are as follows. Firstly, the atoms with similar valence shells are grouped and placed in the same table in increasing order of atomic number $Z$.
A group with  similar ground configurations is divided into two sub-groups according to condition $I\lessgtr6.80$ eV. For atoms with $I>6.80$ eV, the Ps-formation channel is closed. The closest decay channel will be e$^++$A. On the other hand, for $I<6.80$ eV the lowest channel is Ps$+$A$^+$. In this work, the positron binding energies have always been obtained with respect to the decay channel e$^++$A while in many papers the binding energies have been reported relative to the closest decay channel. 
The numbers are related by 
\begin{equation}
  \varepsilon_b = \varepsilon_{Ps} - I + 6.80 \ {\rm eV},
\label{eb-ePs}
\end{equation}
where $\varepsilon_b$ is the positron binding energy relative  to the channel  e$^++$A, $\varepsilon_{Ps}$ is the positron binding energy relative to the channel Ps$+$A$^+$, $I$ is atomic ionization potential, and 6.80 eV is the binding energy of positronium.

The best calculations in the literature are presented in last columns of Tables \ref{t:1} and \ref{t:4}. There are eleven atoms studied accurately to predict  the positron binding energies.

\subsection{Closed-shell atoms} \label{sec:closed}

The results for positron bound with closed-subshell atoms are shown in Table \ref{t:1}. We use these atoms to test our method because these are the easiest systems from the computational point of view and a number of accurate calculations is available.
The positron binding energies of $356$ meV, $514$ meV, $178$ meV, and $103$ meV for Sr, Ca, Cd, and Zn atoms relative to the lowest thresholds (A$^++$Ps for Sr and Ca and e$^++$A for Cd and Zn) have been obtained by the CI$_\infty$FC$_3$ method in Refs. \cite{BM06}, \cite{BM10}, and \cite{MZB08} by Mitroy and co-workers. Here FC$_3$ means fixed core with $3$ particles treated explicitly, CI is the configuration interaction, $\infty$ indicates an extrapolation to $l_{\rm max}\rightarrow \infty$ in the basis expansion. 
All binding energies in the table are presented relative to the positron detachment threshold e$^++$A. Eq. (\ref{eb-ePs}) is used to convert the numbers when needed.

A stochastic variational method (SVM) was used for atoms with a small number of electrons. The positron binding energies of $464$ meV and $86$ meV were obtained for Mg and Be by the SVMFC$_3$ and SVM methods in Refs. \cite{BM06} and \cite{M10}. The most accurate value is probably for Be atom because it is the simplest system (four electrons in closed shells plus  positron).
Our calculations give the binding energy of 214 meV which is 128 meV larger than the 86 meV energy obtained in \cite{M10}. 
A somewhat similar binding energy
access is observed when comparing our results with available accurate calculations for other systems, see Table \ref{t:1} ( we will use this  $128$ meV correction to improve our results for a number of atoms where other accurate calculations are not available - see below).
 Note that the difference between our and the best earlier calculations for heavier  atoms ( Mg, Ca, Zn, Sr, and Cd) could be slightly larger due to the relativistic effects which have not been taken into account in the works by Mitroy group. In contrast, our calculations are relativistic.

Our final raw results for the binding energy ($ \varepsilon_b$) is  the sum of the solution of the eigenvalue equation (\ref{eq:sx}) (SD) and the third order contributions (E3). The negative $ \varepsilon_b < 0 $ means that there is no positron binding to an atom. The positron-atom binding energy is very sensitive to the correlations. This leads  to a large uncertainty in the calculations. Therefore, some negative values of $ \varepsilon_b$ might be within theoretical error bars. 

As expected, increase of $I$ and  decrease of $\alpha$ lead to  decrease of the  positron-atom binding energy $ \varepsilon_b$ (see Table \ref{t:1}). The first subgroup has larger $ \varepsilon_b$ than the rest of the table. This is also expected since  atoms in this subgroup have larger potential strength $S$.

\subsection{Open-shell atoms} \label{sec:open}

Tables \ref{t:2}, \ref{t:3} and \ref{t:4} show the positron binding energies to the ground state configurations of the open-shell atoms, which were suggested in our previous works~\cite{DFG10,DFGH12} as good candidates for experimental detection of positron-atom bound states via resonant annihilation or scattering.

To deal with the positron binding to open-shell atoms with the SD approach we use an approximation in which open shells are treated as closed ones but with
fractional occupation numbers~\cite{DFGH12}. For example, the ground-state electron configuration of neutral Fe is $3d^64s^2$ above the Ar-like core. We treat it as a closed-shell system but reduce the contribution of the $3d$ subshell to the potential and CI matrix elements (\ref{eq:sigma}) by the factor $6/10$. Both members of the fine-structure multiplet, $3d_{3/2}$ and $3d_{5/2}$, are included and  corresponding terms are rescaled by the same factor (see Ref.~\cite{DFGH12} for more details). 
Note that the positron-atom binding has no strong sensitivity to the valence shell being open or closed because the Pauli principle is not applicable to the positron-electron interaction.

Table \ref{t:2} shows the results of our calculations of positron binding
energies $ \varepsilon_b$ for atoms with an open $d$-shell and an open $f$-shell. Their common feature is that they have  an $s$  orbital as the upper subshell  in their ground configurations. Our numerical calculations have shown that  this upper $s$-shell provides dominating contribution to the positron binding. As a result  the positron is bound relative to the threshold of the channel  (A+$e^+$) for all atoms. The situation is different for atoms in Table \ref{t:3} where atoms with an open $p$-shell are presented. Here only In and Tl  have positron binding on their ground state configuration relative to the  channel  (A+$e^+$). 
However,  In and Tl are still unstable against  Ps-formation, see Table \ref{t:5}. All other atoms in Table \ref{t:3}  have no bound states with positron. These results are also consistent with the magnitude of the strength parameter $S$ which is  $S\gtrsim 2.5$ in Table \ref{t:2} and $S\lesssim 2.0$ in Table \ref{t:3}. 

\begin{figure*}
\centering
\epsfig{figure=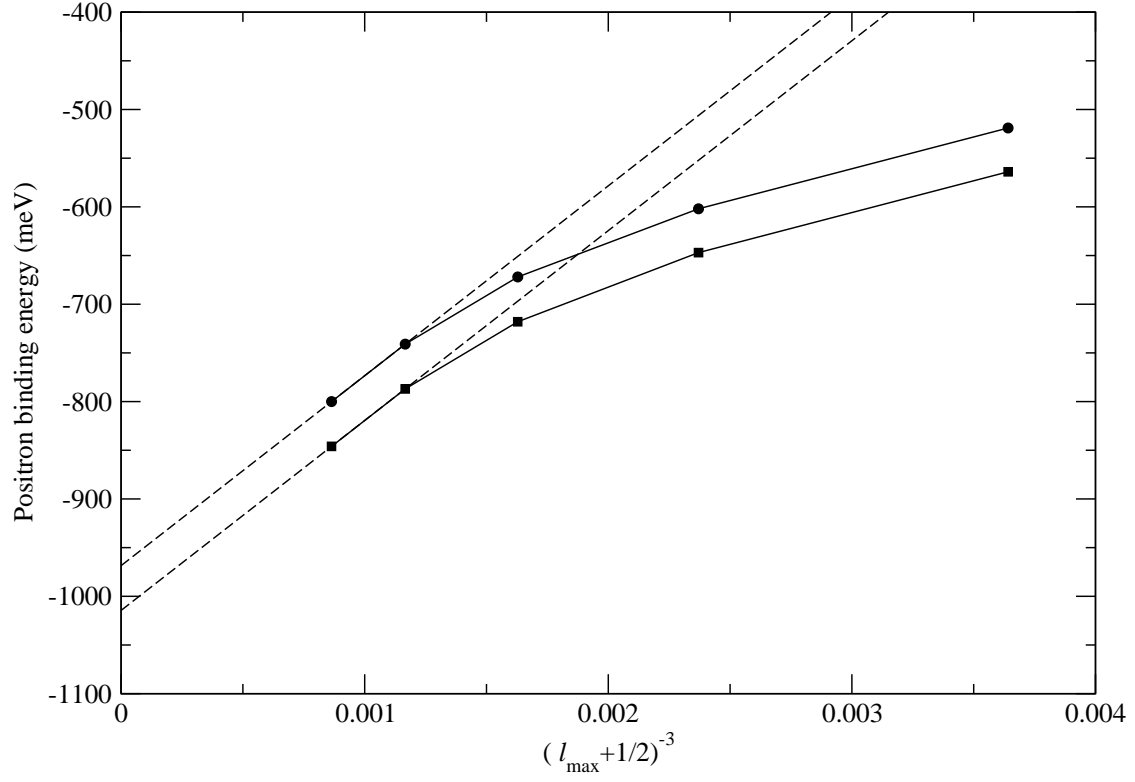,scale=0.6}
\caption{Positron-Li binding energies relative to the channel e$^++$A calculated using the SD equations (black dots) and  with addition of  the third-order correction (black squares) up to $l_{\rm max}=10$. The dashed straight lines mark the extrapolated values on the energy axis for $l_{\rm max}\rightarrow \infty$.} 
\label{f:1}
\end{figure*}

Atoms in Table \ref{t:4} are different  from the rest of the periodic table. First of all, they have very simple electronic configurations, one $s$ electron above closed subshells, so they are the second simplest systems from the  computational point of view  after the closed shell atoms in Table \ref{t:1}. Two particles (an electron plus a positron) above closed shells can be treated using  a sophisticated atomic many-body theory. Some best calculations are presented in the last column of the table. On the other hand, the Ps formation channel is open  for alkali metals (the first set of atoms in Table \ref{t:4}) since  $I<6.80$ eV for them.   
Therefore, these systems are better described as positronium orbiting the positive ion A$^+$ \cite{MBR99} (a molecular type of bonding). Such systems are hard to describe in our present approach where a single-center basis with the origin on the atomic nucleus is used. To achieve a convergence, a very large number of partial waves $l_{\rm max}$ must be included to describe the total wave function. The fact that we could not get any convergent values for the  positron binding energies to Rb, Fr,and Cs supports this argument. Comparison of our result for Li with the accurate variational calculation of Mitroy \cite{M04} shows that our method  underestimates the positron binding energies for alkali atoms. However, we obtained good agreements with the previous accurate calculations for Cu, Ag, and Au ( see Table \ref{t:4}). Therefore,  accurate description for alkali atoms  requires higher values of the maximum angular momentum $l_{\rm max}$. The value $l_{\rm max}=10$ has been fixed in our computation for all atoms. Hence it is reasonable to estimate the extrapolated positron binding energies for alkalis in the limit of $l_{\rm max} \rightarrow \infty$. This is done by using of the asymptotic formula 
\begin{equation}
\varepsilon_b= \varepsilon_b (l_{\rm max})-\frac{A}{(l_{\rm max}+1/2)^3},
\end{equation}    
derived in Ref. \cite{GL02} in the framework of the perturbation theory. Here $\varepsilon_b (l_{\rm max})$ is the binding energy for $l_{\rm max}$ and $A$ is a constant which is different in different atoms. When the convergence is achieved for a given $l_{\rm max}$, all $\varepsilon_b (l>l_{\rm max})$ must lie on a straight line with respect to $1/ (l_{\rm max}+1/2)^3$. For instance, the figure \ref{f:1} shows that the positron binding energy for Li has not been convergent yet at $l_{\rm max}=10$. However, we can estimate the extrapolated positron energy by assuming the straight line obtained from the last two points (for $l_{\rm max}=9$ and $l_{\rm max}=10$). We see that our extrapolated binding energies  for Li and Na are still  smaller than the previous accurate calculations shown in Table \ref{t:4}.  

Now we will try to improve our predictions of the positron binding energies for all atoms  based on comparison of our calculations and available accurate calculations for Be, Li, and Cu. Be is the simplest closed-shell atom that can bind positron (see Table \ref{t:1}), Li is the simplest alkaline atom that can bind positron, and Cu has one electron above closed shell core (see Table \ref{t:4}). Each of them belongs to a different type of group in the periodic table. We assume that atoms in the same group or in nearby similar group interact with positron in the same way. The previous presumably accurate calculations of the positron binding to these three atoms (Be, Li, Cu) can be used to estimate the errors in the present calculations.
Our method overestimates the binding energy for Be~\cite{M10}, underestimates it for Li~\cite{M04}, and is in good agreement with previous calculation for Cu \cite{DFG99}.
Using these differences between our results and the most accurate results of other calculations, we derive recommended values for the positron-atom binding energies  
for the whole periodic table in Table \ref{t:5}. The recommended binding energies for the atoms with closed shells or open $f$, $d$ or $p$ shells
are obtained by subtracting $128$ meV from our results presented in Tables \ref{t:1}, \ref{t:2}, and  \ref{t:3}. This is done 
to eliminate the difference between our value of $214$ meV and the accurate result  $86$ meV \cite{M10} for Be atom. The recommended positron binding energy to Mg has been taken from Ref.~\cite{BM06}. We use our previous accurate results for Cu, Ag, and Au obtained by the relativistic CI+MBPT method~\cite{DFG99,DFH00} as recommended values for these systems ($170$ eV, $123$ eV, $-87$ eV respectively). The previous result for Cu is very close to the current calculation ($166$ eV) anyway. 

The binding energies for  Rb, Cs, and Fr are estimated by the linear extrapolation of the values for  Li, Na, and K with respect to the ion (A$^+$) radius which is inversely proportional to the ion ionization potential ($r_+\sim 1/I^+$). In this estimation we have used the literature results for Li and Na and the corrected  result $2400$ meV  for K atom ( our raw number is $2072$ meV is assumed to be underestimated similar to Li and Na).  Note that  all positron binding energies for Li, Na, and K lies on a straight line with respect to  $1/I^+$. Using values of  $1/I^+$ for  the Rb$^+$, Cs$^+$, and Fr$^+$ ions we obtain the extrapolated positron binding energies for them by putting them  on the same line (a linear extrapolation). We conclude that positron systems with K, Rb, Cs, and Fr atoms are unstable due to decay to positronium and positive ion (see Table \ref{t:5}).

\begin{figure*}
\centering
\epsfig{figure=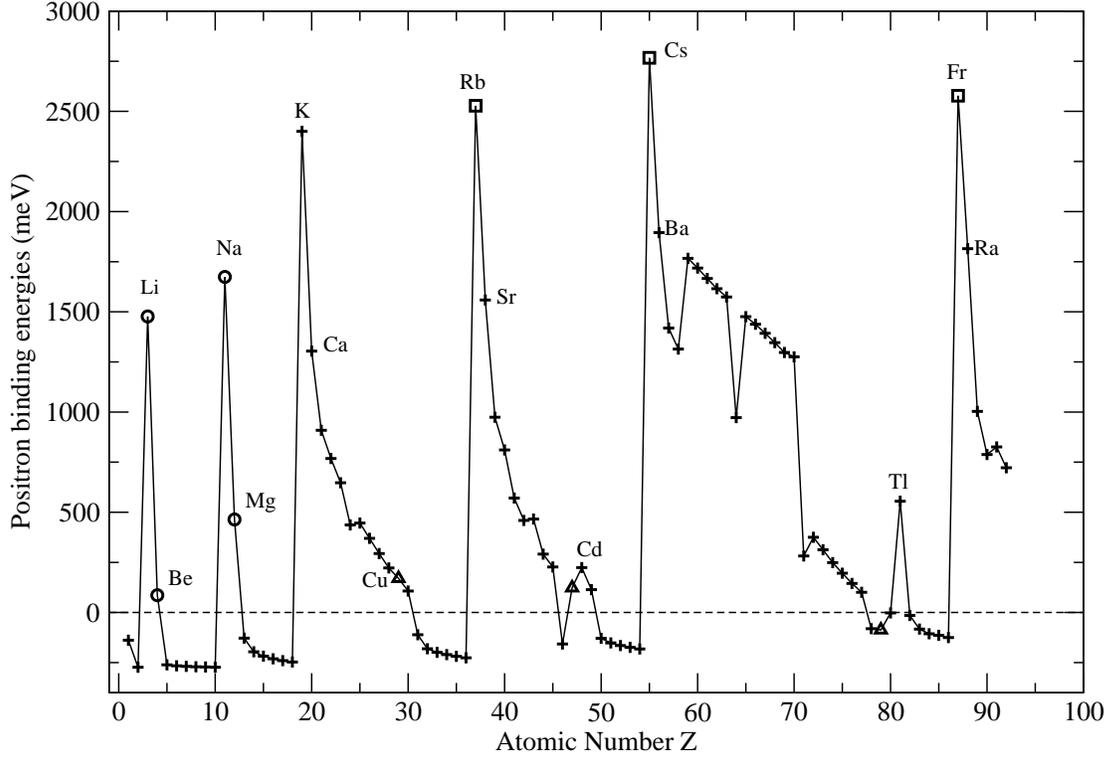,scale=0.6}
\caption{Recommended positron binding energies from Table \ref{t:5} relative  to the dissociation threshold e$^++$A. The results based on current study are shown with $+$ sign. The $\bigcirc$ shows the results of the previous best calculations based on configuration interaction (CI) or stochastic variational methods (SVM). $\bigtriangleup$ shows the previous result of the relativistic method MBPT+CI for Cu, Ag, and Au in our group. The binding energies of Rb, Cs, and Fr are obtained by linear extrapolation of the values of Li, Na, and K with respect to the  ion(A$^+$) radius, which are marked by square $\square$} 
\label{f:2}
\end{figure*}

We also present in Table \ref{t:5}  the positron-atom binding energies relative to the Ps-formation threshold, A$^++$Ps.
 The closest decay channel  is  emphasized by the bold number in the Table \ref{t:5}.  We also present the results  of  the Table \ref{t:5} in the graphical form  on  Fig. \ref{f:2}. 

Therefore, more than a half of atoms in the periodic table may bind positron.

\subsection{Excited states} \label{sec:excited}

In our calculations we do not distinguish between different electron states of the same configuration. If positron is bound to an atom in particular configuration, it is bound to all states of this configuration with approximately the same binding energy. Therefore, the energy splitting inside the ground state configuration is assumed to be the same for atoms with or without a positron.
This assumption is supported by similar features of atomic scalar polarizabilities. The scalar polarizabilities have very close values for the different states of the same ground-state configuration of many-electron atoms~\cite{KDF13}. The values of the scalar polarizabilities  determine the strength of the attractive polarization potential $-\alpha/2r^4$ acting on positron. Therefore,  it is natural to expect that if  the polarizabilities are equal  the  positron binding energies will also be  equal.

 Since the energy splitting within a configuration is assumed to be the same for the atom with or without a positron, the energies of the excited states of the positron-atom system can be obtained as
\begin{equation} \label{eq:rbo}
\varepsilon = E_{\rm ex}- \varepsilon_b,
\end{equation}
where $E_{\rm ex}$ is the experimental value of the atomic excitation energy relative to the ground state  and $\varepsilon_b$ is the positron binding energy to the ground state. These excited states are also bound as long as $\varepsilon<0$. Positive values of $\varepsilon$ correspond to resonances in continuum.  To close the positronium formation channel  we also need the condition
\begin{equation} \label{eq:con}
\varepsilon < I - 6.80 \ {\rm eV}\, ,
\end{equation}
where $I$ is the atomic ionization potential and $6.80$ eV is the binding energy of the Ps  ground state. 

Using condition for the excitation energy $E_{\rm ex} < \varepsilon_b + I - 6.80 \ {\rm eV}$ (stability against the positron emission)  and the recommended positron binding energies $ \varepsilon_b $ from  Table \ref{t:5} we found 
the resonance and bound state energies for 26 atoms. Table \ref{t:6} shows resonances and bound states for the atoms with ionization potential larger than $6.8$ eV while Table \ref{t:7} is for atoms with $I<6.8$ eV.  
Due to the limited accuracy of our calculations  ($\sim 100$ meV) the weakly bound positron states shown in Table \ref{t:6} may turn out to be low-lying resonances and vice versa.

The resonances and bound states in Tables \ref{t:6} and \ref{t:7} have been obtained  for the positron binding to the excited states of the ground configurations. However, positron may bind to a different electron configuration. It is known that helium excited state $1s2s~~^3S$ can bind a positron even though its ground state $1s^2~~^1S$ can not. It has recently been calculated that positron can attach to the $1s2s2p~~^4P^o$ excited state of Li \cite{B12}. Table \ref{t:8} shows our calculations for excited state configurations that can bind  positron relative  to the e$^++$A$^*$ threshold, where A$^*$ is the lowest excited state for a given configuration. 
 We have found that 10 atoms in Table  \ref{t:9} have excited bound states that are stable against both thresholds, e$^++$A$^*$ and A$^{+}+$Ps. The smaller  binding energies are presented  in bold in Table \ref{t:9}. Note that positron does not bind to the ground states of Pd and Pt  but both atoms bind in excited states, see Table \ref{t:9}. 

The energies of the positron resonances ($ \varepsilon >0 $) and excited bound states ($ \varepsilon < 0 $) can be determined from the recommended positron binding energy $ \varepsilon^*_b$ for a particular configuration and excitation energies $E_{\rm ex}$ of the electronic states. Here $E_{\rm ex}$ should also satisfy the condition that $ E_{\rm ex} < \varepsilon^*_b + I - 6.80$ eV  (to stay below Ps formation threshold). 
 The recommended energies of the positron bound states and resonances for excited configurations are presented in Table \ref{t:10}. The results of the present work confirm the claim of Ref.\cite{DFG10} that many open-shell atoms do bind the positron not only in the ground state but also in excited states.

\section{Conclusion} \label{sec:conclusion}

The linearized coupled cluster single-double approach with the third-order correction  is used to calculate the positron binding energy for every atom in the periodic table. The fractional occupation number approximation is used to perform the calculations for open shell atoms. To obtain the recommended values of the positron binding energies we introduce corrections which bring our results in line with the best available calculations which exist for 11 atoms only. We find that 49 atoms can bind positron in the ground state. The recommended values of the binding energies are presented in Table \ref{t:5}. 

   A number of atoms also have excited positron bound states and low energy positron resonances which may be used to measure the positron binding energy in the processes of the positron annihilation and scattering. The recommended values of these excited bound states and resonances are presented in Tables \ref{t:6}, \ref{t:7} and \ref{t:10}. 

An order-of-magnitude estimate of the accuracy of our predictions is $\sim 100$ meV. Due to the limited accuracy some of the calculated weakly bound states may be actually unbound and vice versa.

     Finally, there are two problems for a future study.  If the initial atomic angular momentum $J_A$ is not zero the positron bound states form the doublets with the total angular momenta $J=J_A+1/2$ and $J=J_A-1/2$. The energy splitting is $\sim 1$ meV. 

If $J_A>1/2$ the atom has an electric quadrupole moment $Q$ and produces a long-range potential $$eQ(3 \cos^2\theta-1)/(4r^3)$$
 which decays slower than the polarization potential  $-e^2\alpha/r^4$.  The quadrupole moment is large in atoms with several electrons in an open shell. The quadrupole potential may produce new features in the positron bound states such as localization of the positron wave function in the equatorial or polar areas depending on the sign of $eQ$.  Similar effect may exist for electron in a negative ion.

\begin{acknowledgements}
This work was funded in part by the Australian Research Council. VVF is grateful to the Humboldt Foundation for support and to the Frankfurt Institute for Advanced Studies for hospitality. Authors are grateful to G. F. Gribakin, M. G. Kozlov and D. Budker for helpful discussions. 
\end{acknowledgements}

\begin{table*}
\caption{Positron-atom binding energies relative to the channel e$^++$A ($\varepsilon_b$ in meV) for closed-shell atoms obtained using the SD equations (SD) with the third-order correction (E3). $I$ is the ionization energy from the ground state, $\alpha$ is the polarizability. The combination $S=m\alpha I^2/2 \hslash^2 $ is a dimensionless parameter called the potential strength. The best calculations in the literature are presented in the last column. The negative $\varepsilon_b$ means positron is not bound. Atoms with similar shells in their ground configurations are placed in the same group in increasing order of their atomic numbers $Z$. The first group  is divided into two sub-groups according to $I\lessgtr6.8$ eV.}
\label{t:1}
\begin{ruledtabular}
\begin{tabular}{rll rrrrr rrrr}
\multicolumn{2}{c}{} & 
\multicolumn{1}{c}{ground} &
\multicolumn{1}{c}{$I$} &
\multicolumn{1}{c}{$\alpha_d$\footnotemark[1]} &
\multicolumn{1}{c}{$S$} &
\multicolumn{3}{c}{This work (meV)} &
\multicolumn{1}{c}{Best other} \\
\multicolumn{1}{c}{$Z$} &Atom &
\multicolumn{1}{c}{configuration} &
\multicolumn{1}{c}{(eV)} &
\multicolumn{1}{c}{($10^{-24}$ cm$^3$)} &
\multicolumn{1}{c}{} &
\multicolumn{1}{c}{SD} &
\multicolumn{1}{c}{E3} &
\multicolumn{1}{c}{Total ($\varepsilon_b$) } & 
\multicolumn{1}{c}{cal.(meV)} \\
\hline
20 & Ca & $4s^2$   & 6.113 & 22.8 &3.9& 1382 & 50 & 1432 & 1201 \footnotemark[2] \\ 
38 & Sr & $5s^2$   & 5.695 & 27.6 &4.1& 1638 & 48 & 1687 & 1461\footnotemark[2] \\ 
56 & Ba & $6s^2$  & 5.212 & 39.7 &4.9& 1974 & 48 & 2023 & \\ 
70 & Yb & $4f^{14} 6s^2$ & 6.254 & 20.9 &3.7& 1359 & 43 & 1403 &  \\ 
88 & Ra & $7s^2$   & 5.279 & 38.3 &4.9& 1902 & 40 & 1943 & \\ 
\hline 
2 & He & $1s^2$& 24.587 & 0.205 &0.6& -145 & 0 & -145 &   \\
4 & Be & $2s^2$   & 9.322 & 5.6 &2.2& 187 & 27 & 214 & 86\footnotemark[5]  \\
12 & Mg & $3s^2$   & 7.646 & 10.6 &2.8& 596 & 39 & 636 & 464\footnotemark[3]  \\  
30 & Zn & $4s^2$ & 9.394 & 5.75 &2.3& 211 & 23 & 235 & 103 \footnotemark[6] \\
48 & Cd & $5s^2$ & 8.993 & 7.36 &2.7& 273 & 79 & 352 & 178\footnotemark[4]  \\
80 & Hg & $6s^2$& 10.437 & 5.7 &2.8& 64 & 61 & 126 &  \\ 
\hline
10 & Ne & $2s^22p^6$& 21.564& 0.394& 0.8& -145 & 0 & -145 &    \\
18 & Ar & $3s^23p^6$& 15.759& 1.641& 1.9& -123 & 4 & -119 &    \\
36 & Kr & $4s^24p^6$& 13.999& 2.4844& 2.2& -106 & 8 & -98 &    \\
54 & Xe & $5s^25p^6$& 12.130& 4.044& 2.7& -68 & 14 & -54 &   \\ 
86 & Rn & $6s^26p^6$& 10.748& 5.3& 2.8& -26 & 29 & 3 &  \\
\hline
46 & Pd & $4d^{10}$& 8.34  & 4.8 & 1.5  & -39 & 9 & -29 &  \\
     
\end{tabular}
\footnotetext[1]{Ground-state atomic static dipole polarizabilities from Ref. \cite{Miller}}
\footnotetext[2]{The positron binding energies of 356 meV and 514 meV for atoms Sr and Ca respectively relative  to the lowest threshold A$^++$Ps have been obtained by the CI$_\infty$FC$_3$ method in Ref. \cite{BM06}. Here FC$_3$ means fixed core with $3$ particles treated explicitly, CI is the configuration interaction, $\infty$ indicates an extrapolation to $l_{\rm max}\rightarrow \infty$ in the basis expansion.  In the table, the binding energies relative to the threshold e$^++$A are presented to compare with the present calculations.}
\footnotetext[3]{Calculation by the SVMFC$_3$ method from Ref. \cite{BM06}, where SVM means the stochastic variational method.}
\footnotetext[4]{Calculation by the CI$_\infty$FC$_3$ method from Ref. \cite{BM10}}
\footnotetext[5]{Calculation by the SVM method from Ref. \cite{M10}}
\footnotetext[6]{Calculation by the CI$_\infty$FC$_3$ method from Ref. \cite{MZB08}}
\end{ruledtabular}

\end{table*}

\begin{table*}
\caption{Positron-atom binding energies relative to the channel e$^++$A ($\varepsilon_b$ in meV) for open shell atoms obtained using the SD equations (SD) with third-order correction (E3). $I$ is the ionization energy from the ground state. The combination $S=m\alpha I^2/2\hslash^2$ is called the potential strength. The negative $\varepsilon_b$ means positron is not bound. Atoms with similar shells in their ground configurations are placed in the same group in increasing order of their atomic numbers $Z$. The group with similar ground configurations is divided into two sub-groups according to $I\lessgtr6.8$ eV.}
\label{t:2}
\begin{ruledtabular}
\begin{tabular}{rll rrrrr rrrr}
\multicolumn{2}{c}{} & 
\multicolumn{1}{c}{ground} &
\multicolumn{1}{c}{$I$} &
\multicolumn{1}{c}{$\alpha$\footnotemark[1]} &
\multicolumn{1}{c}{$S$} &
\multicolumn{3}{c}{This work (meV)}  \\
\multicolumn{1}{c}{$Z$} &Atom &
\multicolumn{1}{c}{configuration} &
\multicolumn{1}{c}{(eV)} &
\multicolumn{1}{c}{($10^{-24}$ cm$^3$)} &
\multicolumn{1}{c}{} &
\multicolumn{1}{c}{SD} &
\multicolumn{1}{c}{E3} &
\multicolumn{1}{c}{Total ($\varepsilon_b$) } \\
\hline
21 & Sc & $3d  4s^2$& 6.561& 17.8& 3.5& 908 & 129 & 1037  \\
23 & V  & $3d^34s^2$& 6.746& 12.4& 2.6& 678 & 97 & 775  \\
39 & Y  & $4d  5s^2$& 6.217& 22.7& 4.0& 845 & 256 & 1102   \\
57 & La & $5d  6s^2$& 5.577& 31.1& 4.4& 1223 & 324 & 1547  \\
71 & Lu & $5d  6s^2$& 5.426& 21.9& 2.9& 222 & 245 & 470  \\
89 & Ac & $6d  7s^2$& 5.380& 32.1& 4.2& 706 & 425 & 1131  \\
90 & Th & $6d^27s^2$& 6.307& 32.1& 5.8& 546 & 370 & 916 \\
\hline
22 & Ti & $3d^24s^2$& 6.828& 14.6& 3.1& 785 & 110 & 896  \\
25 & Mn & $3d^54s^2$& 7.435& 9.4& 2.4& 496 & 77 & 574  \\
26 & Fe & $3d^64s^2$& 7.902& 8.4& 2.4& 429 & 69 & 498 \\
27 & Co & $3d^74s^2$& 7.881& 7.5& 2.1& 360 & 61 & 422 \\
28 & Ni & $3d^84s^2$& 7.635& 6.8& 1.8 & 295 & 55 & 350  \\
40 & Zr & $4d^25s^2$& 6.634& 17.9& 3.6& 729 & 209 & 939 \\
43 & Tc & $4d^55s^2$& 7.119& 11.4& 2.6& 461 & 133 & 594 \\

72 & Hf & $5d^26s^2$& 6.825& 16.2& 3.4& 305 & 198 & 503  \\
73 & Ta & $5d^36s^2$& 7.549& 13.1& 3.4& 274 & 166 & 441  \\
74 & W  & $5d^46s^2$& 7.864& 11.1& 3.1& 235 & 141 & 377  \\
75 & Re & $5d^56s^2$& 7.834& 9.7& 2.7& 202 & 121 & 324  \\
76 & Os & $5d^66s^2$& 8.438& 8.5& 2.8& 167 & 105 & 273  \\

77 & Ir & $5d^76s^2$& 8.967& 7.6& 2.8& 137 & 91 & 229 \\

\hline
59 & Pr & $4f^36s^2$& 5.473& 28.2& 3.8& 1786 & 108 & 1895 \\

60 & Nd & $4f^46s^2$& 5.525& 31.4& 4.4& 1746 & 100 & 1846 \\

61 & Pm & $4f^56s^2$& 5.582& 30.1& 4.3& 1701 & 93 & 1794  \\
62 & Sm & $4f^66s^2$& 5.644& 28.8& 4.2& 1655 & 88 & 1743   \\
63 & Eu & $4f^76s^2$& 5.67& 27.7& 4.1& 1617 & 85 & 1702   \\
65 & Tb & $4f^96s^2$& 5.864& 25.5& 4.0& 1525 & 79 & 1604   \\

66 & Dy & $4f^{10}6s^2$&5.939 & 24.5& 3.9& 1490 & 76 & 1566 \\

67 & Ho & $4f^{11}6s^2$& 6.02& 23.6& 3.9& 1446 & 74 & 1521 \\
68 & Er & $4f^{12}6s^2$& 6.107& 22.7& 3.9& 1401 & 72 & 1474 \\
69 & Tm & $4f^{13}6s^2$& 6.184& 21.8& 3.8& 1354 & 71 & 1425 \\
\hline
58 & Ce & $4f  5d6s^2$& 5.538& 29.6& 4.1& 1189 & 253 & 1442 \\
64 & Gd & $4f^75d6s^2$& 6.149& 23.5& 4.0&  816 & 284 & 1100 \\
91 & Pa & $5f^26d7s^2$& 5.89& 25.4& 4.0& 614 & 340 & 954  \\
92 & U  & $5f^36d7s^2$& 6.194& 24.9& 4.4& 517 & 333 & 850 \\
\hline
24 & Cr & $3d^54s$& 6.766& 11.6& 2.4& 488  & 77 & 565   \\

41 & Nb & $4d^45s$& 6.759& 15.7& 3.3& 527 & 172 & 699  \\

42 & Mo & $4d^55s$& 7.092& 12.8& 2.9& 442 & 145 & 587   \\

44 & Ru & $4d^75s$& 7.36& 9.6& 2.4& 310 & 109 & 419  \\

45 & Rh & $4d^85s$& 7.46& 8.6& 2.2& 260 & 95 & 355  \\
78 & Pt & $5d^96s$& 8.959& 6.5& 2.4& -10 & 57 & 47 
\end{tabular}
\footnotetext[1]{Ground-state atomic static dipole polarizabilities from Ref. \cite{Miller}}
\end{ruledtabular}
\end{table*}

\begin{table*}
\caption{Positron-atom binding energies relative to the channel e$^++$A  ($\varepsilon_b$ in meV) for open $p$ shell  obtained using the SD equations (SD) with the third-order correction (E3). $I$ is the ionization energy from the ground state. The combination $S=m\alpha I^2/2\hslash^2$ is the potential strength. The negative $\varepsilon_b$ means positron is not bound. Atoms with similar shells in their ground configurations are placed in the same group in increasing order of their atomic numbers $Z$. The group with similar ground configurations is divided into two sub-groups according to $I\lessgtr6.8$ eV.}
\label{t:3}
\begin{ruledtabular}
\begin{tabular}{rll rrrrr rrrr}
\multicolumn{2}{c}{} & 
\multicolumn{1}{c}{ground} &
\multicolumn{1}{c}{$I$} &
\multicolumn{1}{c}{$\alpha$\footnotemark[1]} &
\multicolumn{1}{c}{$S$} &
\multicolumn{3}{c}{This work (meV)} \\
\multicolumn{1}{c}{$Z$} &Atom &
\multicolumn{1}{c}{configuration} &
\multicolumn{1}{c}{(eV)} &
\multicolumn{1}{c}{($10^{-24}$ cm$^3$)} &
\multicolumn{1}{c}{} &
\multicolumn{1}{c}{SD} &
\multicolumn{1}{c}{E3} &
\multicolumn{1}{c}{Total ($\varepsilon_b$) } \\
\hline
13 & Al & $3s^23p$& 5.986& 6.8 & 1.1& -38 & 38 & 0   \\
31 & Ga & $4s^24p$& 5.999& 8.12& 1.3& -25 & 42 & 17    \\
49 & In & $5s^25p$& 5.786& 10.2& 1.6& 166 & 75 & 242     \\
81 & Tl & $6s^26p$& 6.108& 7.6& 1.3& 588 & 95 & 683    \\
\hline
5 & B & $2s^22p$& 8.298& 3.03& 1.0& -136 & 3 &-133   \\
6 & C & $2s^22p^2$& 11.260& 1.67& 1.0& -139 & 1 & -138  \\
7 & N & $2s^22p^3$& 14.534& 1.10& 1.1& -141 & 0 & -141 \\
8 & O & $2s^22p^4$& 13.618& 0.802& 0.7& -143 & 0 & -143  \\

9 & F & $2s^22p^5$& 17.422& 0.557& 0.8& -144 & 0 & -144 \\
14 & Si & $3s^23p^2$& 8.151& 5.53& 1.7& -88 & 20 & -68    \\
15 & P  & $3s^23p^3$& 10.486& 3.63& 1.8& -102 & 12 & -90    \\
16 & S  & $3s^23p^4$& 10.360& 2.90& 1.4& -111 & 8 & -103    \\
17 & Cl & $3s^23p^5$& 12.967& 2.18& 1.7& -118 & 6 & -112   \\
32 & Ge & $4s^24p^2$& 7.899& 5.84& 1.7& -77 & 24 & -53    \\
33 & As & $4s^24p^3$& 9.789& 4.31& 1.9& -88 & 17 & -71  \\
34 & Se & $4s^24p^4$& 9.752& 3.77& 1.6& -95 & 13 & -82    \\
35 & Br & $4s^24p^5$& 11.814& 3.05& 1.9& -101 & 10 & -91   \\
50 & Sn & $5s^25p^2$& 7.344& 7.84& 1.9& -33 & 32 & -1    \\
51 & Sb & $5s^25p^3$& 8.608& 6.6& 2.2& -49 & 25 & -24     \\
52 & Te & $5s^25p^4$& 9.009& 5.5& 2.0& -57 & 20 & -37     \\
53 & I  & $5s^25p^5$& 10.451& 4.7& 2.3& -63 & 17 & -46  \\
82 & Pb & $6s^26p^2$& 7.416& 6.98& 1.7& 51 & 62 & 113     \\
83 & Bi & $6s^26p^3$& 7.285& 7.4& 1.8& -2 & 46 & 45     \\
84 & Po & $6s^26p^4$& 8.414& 6.8& 2.2& -16 & 38 & 22    \\
85 & At & $6s^26p^5$& 9.350& 6.0& 2.4& -19 & 33 & 14  \\   
\end{tabular}
\footnotetext[1]{Ground-state atomic static dipole polarizabilities from Ref. \cite{Miller}.}
\end{ruledtabular}
\end{table*}

\begin{table*}
\caption{Positron-atom binding energies relative to the channel e$^++$A ($\varepsilon_b$ in meV) for open $s$ shell atoms obtained using the SD equations (SD) with the third-order correction (E3). $I$ is the ionization energy from the ground state. The combination $S=m\alpha I^2/2\hslash^2$ is the potential strength.
The negative $\varepsilon_b$ means positron is not bound, - means the iteration of the SD equations does not converge for those atoms. Atoms with similar shells in their ground configurations are placed in the same group in increasing order of their atomic numbers $Z$. The group with similar ground configurations is divided into two sub-groups according to $I\lessgtr6.8$ eV.}
\label{t:4}
\begin{ruledtabular}
\begin{tabular}{rll rrrrr rrrr}
\multicolumn{2}{c}{} & 
\multicolumn{1}{c}{valence} &
\multicolumn{1}{c}{$I$} &
\multicolumn{1}{c}{$\alpha$\footnotemark[1]} &
\multicolumn{1}{c}{$S$} &
\multicolumn{3}{c}{This work (meV)} &
\multicolumn{1}{c}{Best other} \\
\multicolumn{1}{c}{$Z$} &Atom &
\multicolumn{1}{c}{configuration} &
\multicolumn{1}{c}{(eV)} &
\multicolumn{1}{c}{($10^{-24}$ cm$^3$)} &
\multicolumn{1}{c}{} &
\multicolumn{1}{c}{SD} &
\multicolumn{1}{c}{E3} &
\multicolumn{1}{c}{Total ($\varepsilon_b$) } & 
\multicolumn{1}{c}{cal.(meV)} \\
\hline
3 & Li & $2s$& 5.392& 24.33& 3.2& 800 & 46 & 1015\footnotemark[2] & 1477\footnotemark[4]  \\
11 & Na  & $3s$& 5.139& 24.11 & 2.9 & 1042 & 48 & 1304\footnotemark[2] & 1674\footnotemark[3]   \\
19 & K & $4s$& 4.341& 43.06& 3.7& 1746 & 72 & 2072\footnotemark[2] &  \\
37 & Rb & $5s$& 4.177& 47.24& 3.8& - & - & - &    \\
55 & Cs  & $6s$& 3.894& 59.42& 4.1& - & - & - &    \\
87 & Fr & $7s$& 4.073& 47.1& 3.6& - & - & - &    \\
\hline
1 & H & $1s$& 13.598 & 0.667& 0.6& -138 & 0 & -138  & \\ 
29 & Cu & $4s$& 7.726& 6.2& 1.7& 125 & 40 & 166 &170\footnotemark[7], 152\footnotemark[8]  \\
47 & Ag & $5s$& 7.576& 6.78& 1.8& 172 & 75 & 247 & 123\footnotemark[5], 159\footnotemark[6]   \\
79 & Au & $6s$& 9.225& 5.8& 2.2& -25 & 49 & 24 &  -87\footnotemark[5]\\   
\end{tabular}
\footnotetext[1]{Ground-state atomic static dipole polarizabilities from Ref. \cite{Miller}}
\footnotetext[2]{The positron binding energies of alkali atoms are obtained by extrapolating to the values for $l_{\rm max}\rightarrow \infty$ (see, for instance, Fig. \ref{f:1} for the extrapolation for Li).}
\footnotetext[3]{The positron binding energies of 13 meV for Na relative to the lowest threshold Na$^++$Ps is obtained by the SVMFC$_2$ method in Ref. \cite{RMV98}. Here SVM means the stochastic variational method and FC$_2$ means fixed core with $2$ particles treated explicitly. In the table the binding energy relative to the threshold e$^++$Na is shown to compare with the present calculation.}
\footnotetext[4]{The positron binding energies of 68 meV for Li relative to the lowest threshold Li$^++$Ps is obtained by the SVM method in Ref. \cite{M04}. In the table the binding energy relative to the threshold e$^++$Li is shown to compare with the present calculation.}
\footnotetext[5]{Calculation by the CI+MBPT method, which is the  relativistic configuration interaction plus many-body perturbation theory, from Ref. \cite{DFH00}}
\footnotetext[6]{Calculation by the SVMFC$_2$ method from Ref. \cite{RM98, MBR01}}
\footnotetext[7]{Calculation by the CI+MBPT method from Ref. \cite{DFG99}}
\footnotetext[8]{Calculation by the SVMFC$_2$ method from Ref. \cite{RM9844, MBR01}}
\end{ruledtabular}
\end{table*}

\begin{table*}
\caption{Recommended positron-atom binding energies ($\varepsilon_b$ in meV) for all atoms in the periodic table up to uranium based on the current calculation and the most accurate values in the literature. The negative $ \varepsilon_b$ relative to any threshold shows that positron is not bound. Binding energies relative  to the lowest dissociation threshold are shown in bold. A graphical  presentation of this Table is on  Figure \ref{f:2}. Ionization potentials $I$ are presented in Tables \ref{t:1}$-$\ref{t:4}.}
\label{t:5}
\begin{ruledtabular}
\begin{tabular}{rll rrrrr r}
\multicolumn{2}{c}{} & 
\multicolumn{2}{c}{$\varepsilon_b$(meV) to thresholds } &
\multicolumn{2}{c}{} &
\multicolumn{2}{c}{$\varepsilon_b$(meV) to thresholds} & \\
\multicolumn{1}{c}{$Z$} & Atom &
\multicolumn{1}{c}{e$^+$+A} &
\multicolumn{1}{c}{Ps+A$^+$} &
\multicolumn{1}{c}{$Z$} &
\multicolumn{1}{c}{Atom} &
\multicolumn{1}{c}{e$^+$+A} &
\multicolumn{1}{c}{Ps+A$^+$} &  \\
\hline
 1 & H  & -138     & 6660     & 47 & Ag & {\bf 123}\footnotemark[4]& 899 \\ 
 2 & He & -273     & 17514    & 48 & Cd & {\bf 224} & 2417     \\
 3 & Li & 1477     & {\bf 68} & 49 & In & 114       & -900     \\
 4 & Be & {\bf 86}\footnotemark[1] & 2608     & 50 & Sn & -129 & 415   \\
 5 & B  & -261     & 1237     & 51 & Sb & -152      & 1656     \\
 6 & C  & -266     & 4194     & 52 & Te & -165      & 2044     \\
 7 & N  & -269     & 7465     & 53 & I  & -174      & 3477     \\
 8 & O  & -271     & 6547     & 54 & Xe & -182      & 5148     \\
 9 & F  & -272     & 10350    & 55 & Cs & 2767      & -139  \\
10 & Ne & -273     & 14491    & 56 & Ba & 1895      & {\bf 307}\\ 
11 & Na & 1674     & {\bf 13} & 57 & La & 1419      & {\bf 196}\\ 
12 & Mg & {\bf 464}\footnotemark[2]& 1310     & 58 & Ce & 1314& {\bf 52}\\
13 & Al & -128     & -942     & 59 & Pr & 1767      & {\bf 440}\\
14 & Si & -196     & 1155     & 60 & Nd & 1718      & {\bf 443}\\
15 & P  & -218     & 3468     & 61 & Pm & 1666      & {\bf 448}\\
16 & S  & -231     & 3329     & 62 & Sm & 1615      & {\bf 459}\\
17 & Cl & -240     & 5927     & 63 & Eu & 1574      & {\bf 444}\\
18 & Ar & -247     & 8712     & 64 & Gd & 972       & {\bf 321}\\
19 & K  & 2400     & -59     & 65 & Tb & 1476       & {\bf 540}\\
20 & Ca & 1304     & {\bf 617}& 66 & Dy & 1438      & {\bf 577}\\
21 & Sc & 909      & {\bf 670}& 67 & Ho & 1393      & {\bf 613}\\ 
22 & Ti & {\bf 768}& 796      & 68 & Er & 1346      & {\bf 653}\\
23 & V  & 647      & {\bf 593}& 69 & Tm & 1297      & {\bf 681}\\
24 & Cr & 437      & {\bf 403}& 70 & Yb & 1275      & {\bf 729}\\
25 & Mn & {\bf 446}& 1081     & 71 & Lu & 282       & -1092    \\
26 & Fe & {\bf 370}& 1472     & 72 & Hf & {\bf 375} & 400      \\
27 & Co & {\bf 294}& 1375     & 73 & Ta & {\bf 313} & 1062     \\
28 & Ni & {\bf 222}& 1057     & 74 & W  & {\bf 249} & 1313     \\
29 & Cu & {\bf 170}\footnotemark[3]& 1092 & 75 & Re & {\bf 196} & 1230\\
30 & Zn & {\bf 107}& 2701     & 76 & Os & {\bf 145} & 1783     \\
31 & Ga & -111     & -912     & 77 & Ir & {\bf 101} & 2268     \\ 
32 & Ge & -181     & 918      & 78 & Pt & -81       & 2078     \\
33 & As & -199     & 2790     & 79 & Au & -87\footnotemark[4]& 2400\\
34 & Se & -210     & 2742     & 80 & Hg & -2        & 3635     \\
35 & Br & -219     & 4795     & 81 & Tl & 555       & -137     \\
36 & Kr & -226     & 6973     & 82 & Pb & -15       & 601      \\
37 & Rb & 2528     & -95      & 83 & Bi & -83       & 402      \\
38 & Sr & 1559     & {\bf 454}& 84 & Po & -106      & 1508     \\
39 & Y  & 974      & {\bf 391}& 85 & At & -114      & 2436     \\
40 & Zr & 811      & {\bf 645}& 86 & Rn & -125      & 3823     \\
41 & Nb & 571      & {\bf 530}& 87 & Fr & 2578      & -149  \\ 
42 & Mo & {\bf 459}& 751      & 88 & Ra & 1815      & {\bf 294}\\
43 & Tc & {\bf 466}& 785      & 89 & Ac & 1003      & -417     \\
44 & Ru & {\bf 291}& 851      & 90 & Th & 788       & {\bf 295}\\
45 & Rh & {\bf 227}& 887      & 91 & Pa & 826       & -84      \\
46 & Pd & -157     & 1383     & 92 & U  & 722       & {\bf 116} \\
\end{tabular}
\footnotetext[1]{Calculation by the SVM method from Ref. \cite{M10} is recommended.}
\footnotetext[2]{Calculation by the SVMFC$_3$ method from Ref. \cite{BM06} is recommended.}
\footnotetext[3]{The result of our earlier calculation by the CI+MBPT method (the relativistic configuration interaction plus many-body perturbation theory) from Ref.\cite{DFG99} as a recommended positron binding energy for Cu atom.}
\footnotetext[4]{The results of our earlier calculations by the CI+MBPT method from Ref.\cite{DFH00} as a recommended positron binding energy for Ag and Au atoms.}

\end{ruledtabular}
\end{table*}

\begin{table*}
\caption{Recommended values of excited bound states $ \varepsilon <0 $ or resonance energies ($\varepsilon > 0$) in eV for the measurement of the positron-atom binding energies through resonant annihilation or scattering for atoms which have ionization potential($I$) bigger than $6.8$ eV (the ground state energy of Ps).}
\label{t:6}
\begin{ruledtabular}
\begin{tabular}{rcll llll}
\multicolumn{2}{c}{ }& 
\multicolumn{1}{c}{ground} &
\multicolumn{1}{c}{$I$} &
\multicolumn{1}{c}{Excited states} &
\multicolumn{1}{c}{$E_{\rm ex}$\footnotemark[1]} &
\multicolumn{1}{c}{$ \varepsilon_{b}$} &
\multicolumn{1}{c}{$\varepsilon=E_{\rm ex}-\varepsilon_{b}$} \\
\multicolumn{1}{c}{$Z$} &{Atom} &
\multicolumn{1}{c}{configuration} &(eV) & &(eV)&(eV)& \\
\hline
26 & Fe & $3d^6 4s^2$ &7.902& $^5$D$_{3}$ & 0.052 & 0.370 & $-0.318$ \\ 
   &    &             && $^5$D$_{2}$ & 0.087 & 0.370 & $-0.283$  \\
   &    &             && $^5$D$_{1}$ & 0.110 & 0.370 & $-0.26$ \\
   &    &             && $^5$D$_{0}$ & 0.121 & 0.370 & $-0.249$ \\

27 & Co & $3d^7 4s^2$ &7.881& $^4$F$_{7/2}$ & 0.101 & 0.294 & $-0.193$ \\      
   &    &             && $^4$F$_{5/2}$ & 0.174 & 0.294 & $-0.120$ \\
   &    &             && $^4$F$_{3/2}$ & 0.224 & 0.294 & $-0.070$ \\

28 & Ni & $3d^8 4s^2$ &7.635& $^3$F$_{3}$   & 0.165 & 0.222 &  $-0.057$ \\
   &    &             && $^3$F$_{2}$   & 0.275 & 0.222 &  0.053  \\
   
44 & Ru & $4d^7 5s$   &7.36& $^5$F$_{4}$   & 0.148 & 0.291 & $-0.143$ \\
   &    &             && $^5$F$_{3}$   & 0.259 & 0.291 & $-0.032$ \\
   &    &             && $^5$F$_{2}$   & 0.336 & 0.291 & 0.045    \\
   &    &             && $^5$F$_{1}$   & 0.385 & 0.291 & 0.094    \\
   &    &             && $^3$F$_{4}$   & 0.811 & 0.291 & 0.520 \\
 
45 & Rh & $4d^8 5s$   &7.46& $^4$F$_{7/2}$ & 0.190 & 0.227 & $-0.037$ \\ 
   &    &             && $^4$F$_{5/2}$ & 0.322 & 0.227 &  0.095 \\
   &    &             && $^4$F$_{3/2}$ & 0.430 & 0.227 &  0.203 \\
   &    &             && $^2$F$_{7/2}$ & 0.706 & 0.227 & 0.479 \\
   
72 & Hf & $5d^2 6s^2$ &6.825& $^3$F$_{3}$   & 0.292 & 0.375 & $-0.083$ \\

73 & Ta & $5d^3 6s^2$ &7.549& $^4$F$_{5/2}$ & 0.249 & 0.313 & $-0.064$ \\ 
   &    &             && $^4$F$_{7/2}$ & 0.491 & 0.313 & 0.178 \\
   &    &             && $^4$F$_{9/2}$ & 0.697 & 0.313 & 0.384 \\
   &    &             && $^4$P$_{1/2}$ & 0.750 & 0.313 & 0.437 \\
   &    &             && $^4$P$_{3/2}$ & 0.752 & 0.313 & 0.439 \\
   
74 & W  & $5d^4 6s^2$ &7.864& $^5$D$_{1}$ & 0.207 & 0.249 & $-0.042 $ \\
   &    &             && $^5$D$_{2}$ & 0.412 & 0.249 & 0.163 \\
   &    &             && $^5$D$_{3}$ & 0.599 & 0.249 & 0.350 \\
   &    &             && $^5$D$_{4}$ & 0.771 & 0.249 & 0.522 \\         
   &    &             && $^3$P2$_{0}$& 1.181 & 0.249 & 0.932 \\

76 & Os & $5d^6 6s^2$ &8.438& $^5$D$_{3}$ & 0.516 & 0.145 & 0.371 \\
   &    &             && $^5$D$_{2}$ & 0.340 & 0.145 & 0.195 \\
   &    &             && $^5$D$_{1}$ & 0.715 & 0.145 & 0.570 \\
   &    &             && $^5$D$_{0}$ & 0.755 & 0.145 & 0.610 \\
   &    &             && $^3$H$_{5}$ & 1.778 & 0.145 & 1.633 \\

77 & Ir & $5d^7 6s^2$ &8.967& $^4$F$_{3/2}$ & 0.506 & 0.101 & 0.405 \\
   &    &             && $^4$F$_{5/2}$ & 0.717 & 0.101 & 0.616 \\
   &    &             && $^4$F$_{7/2}$ & 0.784 & 0.101 & 0.683 \\
   &    &             && $^2$G$_{9/2}$ & 1.728 & 0.101 & 1.627 \\
   &    &             && $^2$G$_{7/2}$ & 2.204 & 0.101 & 2.103 \\
   &    &             && $^4$P$_{5/2}$ & 1.997 & 0.101 & 1.896 \\
   
\end{tabular}
\footnotetext[1]{Atomic excitation energy  relative to
 the ground state from Ref.~\cite{NIST}.}
\end{ruledtabular}
\end{table*}

\begin{table*}
\caption{Recommended energies of the excited positron-atom bound states $ \varepsilon <0 $ below Ps formation threshold ($ \varepsilon<I-6.8$ eV) for atoms which have ionization potential ($I$) less than 6.8 eV (the ground state energy of Ps).}
\label{t:7}
\begin{ruledtabular}
\begin{tabular}{rcll llll}
\multicolumn{2}{c}{ }& 
\multicolumn{1}{c}{ground} &
\multicolumn{1}{c}{$I$} &
\multicolumn{1}{c}{Excited states} &
\multicolumn{1}{c}{$E_{\rm ex}$\footnotemark[1]} &
\multicolumn{1}{c}{$ \varepsilon_{b}$} &
\multicolumn{1}{c}{$ \varepsilon=E_{\rm ex}-\varepsilon_{b}$} \\
\multicolumn{1}{c}{$Z$} &{Atom} &
\multicolumn{1}{c}{configuration} &(eV)& &(eV)&(eV)& \\
\hline
21 & Sc & $3d 4s^2$    &6.561& $^2$D$_{5/2}$ & 0.021 & 0.909 & $-0.888$ \\ 

22 & Ti & $3d^2 4s^2$  &6.828& $^3$F$_{3}$ & 0.021 & 0.768 & $-0.747$ \\
   &   &               && $^3$F$_{4}$ & 0.048 & 0.768 & $-0.720$ \\
     
23 & V & $3d^3 4s^2$   &6.746& $^4$F$_{5/2}$ & 0.017 & 0.647 & $-0.63$ \\
   &   &               && $^4$F$_{7/2}$ & 0.040 & 0.647 & $-0.607$ \\
   &   &               && $^4$F$_{9/2}$ & 0.068 & 0.647 & $-0.579$ \\
         
39 & Y & $4d 5s^2$     &6.217& $^2$D$_{5/2}$ & 0.066 & 0.974 & $-0.908$ \\

40 & Zr & $4d^2 5s^2$ &6.634& $^3$F$_{3}$   & 0.071 & 0.811 & $-0.74$ \\
   &    &             && $^3$F$_{4}$   & 0.154 & 0.811 & $-0.657$ \\ 
   &    &             && $^3$P$_{2}$   & 0.519 & 0.811 & $-0.292$ \\
   &    &             && $^3$P$_{0}$   & 0.520 & 0.811 & $-0.291$ \\
   &    &             && $^3$P$_{1}$   & 0.542 & 0.811 & $-0.269$ \\
   &    &             && $^1$D$_{2}$   & 0.632 & 0.811 & $-0.179$ \\   

41 & Nb & $4d^4 5s $   &6.759& $^6$D$_{3/2}$  & 0.019 & 0.571 & $-0.552$ \\
   &    &              && $^6$D$_{5/2}$ & 0.049 & 0.571 & $-0.522$ \\
   &    &              && $^6$D$_{7/2}$ & 0.086 & 0.571 & $-0.485$ \\
   &    &              && $^6$D$_{9/2}$ & 0.130 & 0.571 & $-0.441$ \\

57 & La & $5d 6s^2$    &5.577& $^2$D$_{5/2}$ & 0.130 & 1.419 & $-1.289$ \\

58 & Ce & $4f5d 6s^2$    &5.538& $^3$F$_{2}^o$& 0.028 & 1.314 & $-1.286$ \\

59 & Pr & $4f^3 6s^2$  &5.473& $^4$I$_{11/2}^o$& 0.171& 1.767 & $-1.596$ \\
   &    &              && $^4$I$_{13/2}^o$& 0.353 & 1.767 & $-1.414$ \\

60 & Nd & $4f^4 6s^2$   &5.525& $^5$I$_{5}$ & 0.140 & 1.718 &$-1.578$ \\ 
   &    &               && $^5$I$_{6}$ & 0.293 & 1.718 &$-1.425$ \\

61 & Pm & $4f^5 6s^2$ &5.582& $^6$H$^o_{7/2}$ & 0.100 & 1.666& $-1.566$  \\
   &    &             && $^6$H$^o_{9/2}$ & 0.217 & 1.666 & $-1.449$  \\
   &    &             && $^6$H$^o_{11/2}$& 0.347 & 1.666 & $-1.319$  \\

62 & Sm & $4f^6 6s^2$ &5.644& $^7$F$_{1}$ & 0.036 & 1.615 & $-1.579$  \\ 
   &    &             && $^7$F$_{2}$ & 0.101 & 1.615 & $-1.514$  \\
   &    &             && $^7$F$_{3}$ & 0.185 & 1.615 & $-1.430$  \\
   &    &             && $^7$F$_{4}$ & 0.282 & 1.615 & $-1.333$  \\
   &    &             && $^7$F$_{5}$ & 0.388 & 1.615 & $-1.227$  \\  
    
64 & Gd  & $4f^75d6s^2$ &6.149& $^9$D$^o_{3}$ & 0.027 & 0.972 & $-0.945$ \\         
   &     &              && $^9$D$^o_{4}$ & 0.066 & 0.972 & $-0.906$ \\
   &     &              && $^9$D$^o_{5}$ & 0.124 & 0.972 & $-0.848$ \\
   &     &              && $^9$D$^o_{6}$ & 0.213 & 0.972 & $-0.759$ \\

65 & Tb & $4f^9 6s^2$ &5.864& $^6$H$^o_{13/2}$& 0.344& 1.476 & $-1.132$  \\

66 & Dy & $4f^{10} 6s^2$ &5.939& $^5$I$_{7}$ & 0.513 & 1.438 & $-0.925$ \\

 68  & Er   & $4f^{12} 6s^2$ &6.107& $^3$F$_{4}$ & 0.624 & 1.346 & $-0.722$ \\
\end{tabular}
\footnotetext[1]{Atomic excitation energy relative to
 the ground state from Ref.~\cite{NIST}.}
\end{ruledtabular}
\end{table*}

\begin{table*}
\caption{Positron-atom binding energies ($\varepsilon^*_b$ in meV) for the excited states of  configurations different from the ground state configurations which are obtained using the SD equations (SD) with the third-order correction (E3). $E_{\rm ex}$ is the excitation energy relative to the ground state. }
\label{t:8}
\begin{ruledtabular}
\begin{tabular}{rll rrrrr rrrr}
\multicolumn{2}{c}{} & 
\multicolumn{1}{c}{valence} &
\multicolumn{1}{c}{$I$} &
\multicolumn{1}{c}{$E_{\rm ex}$\footnotemark[1]} &
\multicolumn{3}{c}{This work (meV)}  \\
\multicolumn{1}{c}{$Z$} &Atom &
\multicolumn{1}{c}{configuration} &(eV)&
\multicolumn{1}{c}{(eV)} &
\multicolumn{1}{c}{SD} &
\multicolumn{1}{c}{E3} &
\multicolumn{1}{c}{Total ($\varepsilon^*_b$) } \\
\hline
21   & Sc  & $3d^24s $    &6.561& 1.428 & 849  & 109 & 958  \\
22   & Ti  & $3d^34s $    &6.828& 0.813 & 727  & 97  & 825  \\
23   & V   & $3d^44s $    &6.746& 0.262 & 602  & 86  & 689  \\
24   & Cr  & $3d^44s^2$   &6.766& 0.961 & 590  & 86  & 676  \\
25   & Mn  & $3d^64s $    &7.435& 2.114 & 382  & 68  & 450  \\
26   & Fe  & $3d^74s $    &7.902& 0.859 & 315  & 61  & 376  \\
27   & Co  & $3d^84s $    &7.881& 0.432 & 243  & 53  & 297  \\
28   & Ni  & $3d^94s $    &7.635& 0.025 & 173  & 46  & 220  \\
29   & Cu  & $3d^94s^2$   &7.726& 1.389 & 240  & 49  & 289  \\
39   & Y   & $4d^2 5s$    &6.217& 1.356 & 683  & 258 & 942  \\
40   & Zr  & $4d^35s $    &6.634& 0.604 & 623  & 208 & 831  \\
41   & Nb  & $4d^35s^2$   &6.759& 0.141 & 658  & 178 & 836  \\
42   & Mo  & $4d^45s^2$   &7.092& 1.359 & 583  & 155 & 739  \\
43   & Tc  & $4d^65s $    &7.119& 0.518 & 355  & 124 & 479  \\
44   & Ru  & $4d^65s^2$   &7.36& 0.927 & 461  & 121 & 583  \\
46   & Pd  & $4d^{9}5s$   &8.34& 0.814 & 205  & 83  & 288  \\
     &     & $4d^{8}5s^2$ && 3.112 & 361  & 97  & 459  \\
56   & Ba  & $6s5d$       &5.212& 1.120 & 1345 & 389 & 1734 \\
57   & La  & $5d^26s $    &5.577& 0.331 & 1237 & 324 & 1561 \\
59   & Pr  & $4f^25d6s^2$ &5.473& 0.549 & 1120 & 286 & 1406 \\
60   & Nd  & $4f^35d6s^2$ &5.525& 0.838 & 1060 & 285 & 1345 \\
64   & Gd  & $4f^75d^26s$ &6.149& 0.790 & 544  & 288 & 832  \\
65   & Tb  & $4f^85d6s^2$ &5.864& 0.035 & 749  & 284 & 1033 \\
66   & Dy  & $4f^95d6s^2$ &5.939& 0.938 & 672  & 283 & 955  \\
72   & Hf  & $5d^36s $    &6.825& 1.747 & 349  & 190 & 539  \\
73   & Ta  & $5d^46s $    &7.549& 1.210 & 126  & 147 & 273  \\
74   & W   & $5d^56s $    &7.864& 0.366 & 83   & 118 & 201  \\
75   & Re  & $5d^66s $    &7.834& 1.457 & 51   & 96  & 147  \\
76   & Os  & $5d^76s $    &8.438& 0.638 & 25   & 79  & 105  \\
77   & Ir  & $5d^86s $    &8.967& 0.351 & 5    & 67  & 72   \\
78   & Pt  & $5d^86s^2$   &8.959& 0.102 & 111  & 80  & 191  \\
     &     & $5d^{10}$    && 0.761 & 20   & 23  & 44    
\end{tabular}
\footnotetext[1]{Excitation energies relative to the ground states from Ref. \cite{NIST}}
\end{ruledtabular}
\end{table*}

\begin{table}
\caption{Recommended positron-atom binding energies ($\varepsilon^*_b$ in meV) for excited states of all atoms in Table \ref{t:8}. The negative $ \varepsilon^*_b$ relative  to any threshold shows that positron is not bound. Binding energies relative to the lowest dissociation threshold are shown in bold. Ionization potentials $I$ and electron excitation energies $E_{\rm ex}$  are presented in Table \ref{t:8}.}
\label{t:9}
\begin{ruledtabular}
\begin{tabular}{rll rrrrr r}
\multicolumn{2}{c}{} &
\multicolumn{1}{c}{valence} &
\multicolumn{2}{c}{$\varepsilon^*_b$(meV) to thresholds}  \\
\multicolumn{1}{c}{$Z$} &
\multicolumn{1}{c}{Atom} &
\multicolumn{1}{c}{configuration} &
\multicolumn{1}{c}{e$^+$+A$^\ast$} &
\multicolumn{1}{c}{Ps+A$^{+}$} \\
\hline
21   & Sc  & $3d^24s$      & 830       & -837      \\
22   & Ti  & $3d^34s$      & 697       & -88       \\
23   & V   & $3d^44s$      & 561       & {\bf 245} \\
24   & Cr  & $3d^44s^2$    & 548       & -447      \\
25   & Mn  & $3d^64s $     & 322       & -1157     \\
26   & Fe  & $3d^74s $     & {\bf 248} & 491       \\
27   & Co  & $3d^84s $     & {\bf 169} & 818       \\
28   & Ni  & $3d^94s $     & {\bf 92}  & 902       \\
29   & Cu  & $3d^94s^2$    & 161       & -302      \\
39   & Y   & $4d^2 5s$     & 814       & -1125     \\
40   & Zr  & $4d^35s $     & 703       & -67       \\
41   & Nb  & $4d^35s^2$    & 708       & {\bf 526} \\
42   & Mo  & $4d^45s^2$    & 611       & -456      \\
43   & Tc  & $4d^65s $     & 351       & {\bf 152} \\ 
44   & Ru  & $4d^65s^2$    & 455       & {\bf 88}  \\
46   & Pd  & $4d^{9}5s$    & {\bf 160} & 886       \\
     &     & $4d^{8}5s^2$  & 331       & -1241     \\
56   & Ba  & $6s5d$        & 1606      & -1102     \\
57   & La  & $5d^26s$      & 1433      & -121      \\
59   & Pr  & $4f^25d6s^2$  & 1278      & -598      \\
60   & Nd  & $4f^35d6s^2$  & 1217      & -896      \\
64   & Gd  & $4f^75d^26s$  & 704       & -737      \\
65   & Tb  & $4f^85d6s^2$  & 905       & -66       \\
66   & Dy  & $4f^95d6s^2$  & 827       & -972      \\
72   & Hf  & $5d^36s $     & 411       & -1311     \\
73   & Ta  & $5d^46s $     & 145       & -316      \\
74   & W   & $5d^56s $     & {\bf 73}  & 771       \\
75   & Re  & $5d^66s $     & 19        & -404      \\
76   & Os  & $5d^76s $     & -23       & 977       \\
77   & Ir  & $5d^86s $     & -56       & 1760      \\
78   & Pt  & $5d^86s^2$    & {\bf 63}  & 2120      \\
     &     & $5d^{10}$     & -84       & 1314      \\ 
\end{tabular}
\end{ruledtabular}
\end{table}

\begin{table*}
\caption{Recommended positron-atom resonances $ \varepsilon >0 $ and excited bound states $ \varepsilon < 0 $ below Ps formation threshold ($ \varepsilon<I-6.8$ eV) for atoms which can attach positron in its excited configuration from Table \ref{t:9}. The ground state ionization potentials $I$ are also presented.}
\label{t:10}
\begin{ruledtabular}
\begin{tabular}{rcll llll}
\multicolumn{2}{c}{ }& 
\multicolumn{1}{c}{Valence} &
\multicolumn{1}{c}{$I$} &
\multicolumn{1}{c}{Excited states} &
\multicolumn{1}{c}{$E_{\rm ex}$\footnotemark[1]} &
\multicolumn{1}{c}{$ \varepsilon^\ast_{b}$} &
\multicolumn{1}{c}{$ \varepsilon=E_{\rm ex}-\varepsilon^ \ast _{b}$} \\
\multicolumn{1}{c}{$Z$} &{Atom} &
\multicolumn{1}{c}{configuration} &(eV)& &(eV)&(eV)& \\
\hline
4  &Be & $1s^2 2s 2p$& 9.322& $^3P^o$ &   2.725 & & 2.49\footnotemark[2]\\        
23 & V & $3d^4 4s$  &6.746& $^6$D$_{1/2}$ & 0.262 & 0.561 & $-0.299$ \\
   &   &            && $^6$D$_{3/2}$ & 0.267 & 0.561 & $-0.294$ \\
   &   &            && $^6$D$_{5/2}$ & 0.275 & 0.561 & $-0.286$ \\
   &   &            && $^6$D$_{7/2}$ & 0.286 & 0.561 & $-0.275$ \\
   &   &            && $^6$D$_{9/2}$ & 0.301 & 0.561 & $-0.260$ \\

26 & Fe & $3d^7 4s$    &7.902& $^5$F$_{5}$ & 0.859 & 0.248 & 0.611 \\
   &   &               && $^5$F$_{4}$ & 0.915 & 0.248 & 0.667 \\
   &   &               && $^5$F$_{3}$ & 0.958 & 0.248 & 0.710 \\
   &   &               && $^5$F$_{2}$ & 0.990 & 0.248 & 0.742 \\
   &   &               && $^5$F$_{1}$ & 1.011 & 0.248 & 0.763 \\
 
27 & Co & $3d^8 4s$   &7.881& $^4$F$_{9/2}$ & 0.432 & 0.169 & 0.263 \\
   &    &             && $^4$F$_{7/2}$ & 0.513 & 0.169 & 0.344 \\
   &    &             && $^4$F$_{5/2}$ & 0.581 & 0.169 & 0.412 \\
   &    &             && $^4$F$_{3/2}$ & 0.629 & 0.169 & 0.46    \\
   &    &             && $^2$F$_{7/2}$ & 0.922 & 0.169 & 0.753    \\
   &    &             && $^2$F$_{5/2}$ & 1.049 & 0.169 & 0.88 \\

28 & Ni & $3d^9 4s$   &7.635& $^3$D$_{3}$   & 0.025 & 0.092 &  $-0.067$ \\
   &    &             && $^3$D$_{2}$   & 0.109 & 0.092 &  0.017 \\
   &    &             && $^3$D$_{1}$   & 0.212 & 0.092 &  0.12  \\   
   &    &             && $^1$D$_{2}$   & 0.422 & 0.092 &  0.33  \\

41 & Nb & $4d^3 5s^2$  &6.759& $^4$F$_{3/2}$ & 0.141 & 0.708 & $-0.567$ \\
   &    &              && $^4$F$_{5/2}$ & 0.196 & 0.708 & $-0.512$ \\
   &    &              && $^4$F$_{7/2}$ & 0.267 & 0.708 & $-0.441$ \\
   &    &              && $^4$F$_{9/2}$ & 0.347 & 0.708 & $-0.361$ \\
   &    &              && $^4$P$_{1/2}$ & 0.619 & 0.708 & $-0.089$ \\
   &    &              && $^4$P$_{3/2}$ & 0.656 & 0.708 & $-0.052$ \\

43 & Tc & $4d^6 5s$   &7.119& $^6$D$_{1/2}$ & 0.518 & 0.351 & 0.167 \\ 
   &    &             && $^6$D$_{3/2}$ & 0.496 & 0.351 & 0.145 \\ 
   &    &             && $^6$D$_{5/2}$ & 0.458 & 0.351 & 0.107 \\
   &    &             && $^6$D$_{7/2}$ & 0.403 & 0.351 & 0.052 \\
   &    &             && $^6$D$_{9/2}$ & 0.318 & 0.351 & $-0.033$ \\

44 & Ru & $4d^6 5s^2$   &7.36& $^5$D$_{4}$   & 0.927 & 0.455 & 0.472 \\

46 & Pd & $4d^9 5s$   &8.34& $^2[5/2]_{3}$ & 0.813 & 0.160 & 0.653 \\
   &    &             && $^2[5/2]_{2}$ & 0.961 & 0.160 & 0.801 \\ 
   &    &             && $^2[3/2]_{1}$ & 1.251 & 0.160 & 1.091 \\
   &    &             && $^2[3/2]_{2}$ & 1.453 & 0.160 & 1.293 \\

74 & W  & $5d^5 6s$   &7.864& $^7$S$_{3}$ & 0.365 & 0.073 & 0.292 \\

78 & Pt & $5d^8 6s^2$ &8.959& $^3$F$_{4}$ & 0.102 & 0.063 & 0.039 \\
   &    &             && $^3$F$_{3}$ & 1.254 & 0.063 & 1.191 \\
   &    &             && $^3$F$_{2}$ & 1.921 & 0.063 & 1.858 \\
   &    &             && $^3$P$_{2}$ & 0.814 & 0.063 & 0.751     
\end{tabular}
\footnotetext[1]{Atomic excitation energy relative to
 the ground state from Ref.~\cite{NIST}.}
\footnotetext[2]{The recent calculation of a resonant level from Ref.~\cite{BP13}.}
\end{ruledtabular}
\end{table*}

\end{document}